\documentclass{PoS}

\title{Glueball spectral densities from the lattice}

\ShortTitle{Glueball spectral density}

\author{\speaker{O.~Oliveira} \\ 
        CFC, Departamento de F\'{\i}sica, Universidade de Coimbra, 3004 516 Coimbra, Portugal\\
        E-mail: \email{orlando@teor.fis.uc.pt}}

\author{D.~Dudal \\
        Ghent University, Department of Physics and Astronomy, Krijgslaan 281-S9, 9000 Gent, Belgium\\
        E-mail: \email{david.dudal@ugent.be}}

\author{P.~J.~Silva \\
        CFC, Departamento de F\'{\i}sica, Universidade de Coimbra, 3004 516 Coimbra, Portugal\\
        E-mail: \email{psilva@teor.fis.uc.pt}}

\abstract{ The propagator of a physical degree of freedom ought to obey a K\"{a}ll\'{e}n-Lehmann spectral representation, with positive spectral density. The latter quantity is directly related to a cross section based on the optical theorem. The spectral density is a crucial ingredient of a quantum field theory with elementary and bound states, with a direct experimental connection as the masses of the excitations reflect themselves into (continuum) $\delta$-singularities. In usual lattice simulational approaches to the QCD spectrum the spectral density itself is not accessed. The (bound state) masses are extracted from the asymptotic exponential decay of the two-point function.  Given the importance of the spectral density, each nonperturbative continuum approach to QCD should be able to adequately describe it or to take into proper account. In this work, we wish to present a first trial in extracting an estimate for the scalar glueball spectral density in SU(3) gluodynamics using  lattice gauge theory.
}

\FullConference{The 30 International Symposium on Lattice Field Theory - Lattice 2012,\\
		June 24-29, 2012\\
		Cairns, Australia}

\begin{document}

\section{Introduction and Motivation}

The observed hadronic physical spectra can be understood as a collection of bound states of quarks and gluons,
the fundamental fields of QCD. So far, all attempts to detect directly quarks and gluons have produced negative
results. However, if one takes QCD seriously, one has to explain why quarks and gluons do not contribute to
the S-matrix or, from another point of view, why quarks and gluons only come about when they are
confined into color singlet
states. Investigating the spectral representation of the QCD propagators does not only help us to perceive better
the dynamics of the Yang-Mills theory but it also provides clues to grasp the problem of confinement.

Let $\mathcal{O}(p)$ be the Euclidean momentum creation operator associated with a given set of quantum numbers.
Its vacuum expectation value $\langle \mathcal{O}(p) ~ \mathcal{O}(-p) \rangle$ defines a propagator which
encodes all the information about the quanta with the chosen quantum numbers. Furthermore, let us assume that
the propagator has a K\"all\'en-Lehmann spectral representation, i.e.~that
\begin{equation}
   \langle \mathcal{O}(p) ~ \mathcal{O}(-p) \rangle = \int_0^{+\infty} d \mu ~  \frac{\rho (\mu) }{p^2 + \mu} \ ,
\end{equation}
where $\rho (\mu)$ is the spectral density, which is real and positive defined if the quantum associated with
$\mathcal{O}(p)$ contributes physically to the S-matrix. Then, if the spectral density vanishes or becomes negative
it is a sign that the corresponding quanta are not asymptotic states of the S-matrix. Of course, this is not
a proof of confinement, but it is a necessary condition to be satisfied by a confined particle (assuming that
its propagator has a K\"all\'en-Lehmann representation at least). If the quanta associated with $\mathcal{O}$
contribute to the S-matrix, a typical spectral density is given by
\begin{equation}
  \rho ( \mu ) = \sum^N_{i=1} Z_i \, \delta ( \mu - \mu_i) ~ + ~ \theta ( \mu - \mu_0 ) \, f( \mu ) \, ,
\end{equation}
where $\mu_i$ are the particle masses squared associated with the operator $\mathcal{O}$, $Z_i$ the corresponding
probability of creating such a state and $\mu_0$ is the threshold where
 the multiparticle spectrum (= the continuum) starts.

The lattice approach to QCD provides a way to compute nonperturbatively the propagators of quarks, gluons
and composite operators, such us glueballs, from first principles.
The problem we address here is what can be learned about the spectral density given a propagator.
In particular in this contribution, we will revisit the Landau gauge gluon propagator and
report on preliminary data for the glueball propagator. In the present report we will consider only pure gauge
Yang-Mills SU(3). For the generation of the gauge configurations, we use the MILC code \cite{MILC}.
The simulations were performed both at Coimbra and Ghent Universities.

From the numerical point of view, the computation of $\rho ( \mu )$ from the propagator requires an inverse
integral transform which, typically, calls for some regularization of the data as provided by, for example,
the maximum entropy method \cite{Asakawa:2000tr}.
Spectral forms of correlation functions are widely studied using lattice simulations, in particular mesons,
charmonia \textit{etc}. at finite $T$. Note that,
although it will be interesting to access the propagators in the entire complex $p^2$ plane,
lattice QCD simulations only allow us to consider the Euclidean momenta region $p^2\geq 0$.

\section{On the gluon propagator}

The Landau gauge gluon propagator
\begin{equation}
  D^{ab}_{\mu\nu}Ê(q) = \delta^{ab} \left( \delta_{\mu\nu} - \frac{q_\mu q_\nu}{q^2} \right) D(q^2)
\end{equation}
computed on the lattice has been thoroughly scrutinized in the recent years, see for example
\cite{Dudal:2012zx,Oliveira:2012eh,Cucchieri:2011ig,Dudal:2010tf,Bogolubsky:2009dc,Cucchieri:2007md}. It turns
out that the propagator is well described by the tree-level prediction of the refined Gribov-Zwanziger
action \cite{Dudal:2008sp}
\begin{equation}
   D(q^2) = Z \frac{ q^2 + M^2 }{q^4 + (M^2 + m^2) q^2 + \lambda^4}
   \label{Eq:RGZ}
\end{equation}
up to momenta $ q \approx 4$ GeV with $Z \sim 0.81$, $M^2 \sim 4.0$ GeV$^2$, $M^2 + m^2 \sim 0.57$ GeV$^2$
and $\lambda^4 \sim 0.39$ GeV$^4$. This functional form means that the low energy propagator has two complex
poles at $\approx 0.412 \pm i \, 0.674$ GeV. Given that (\ref{Eq:RGZ}) is a perturbative result it is conceivable that
taking into account quantum corrections will modify the observed behaviour, leading to branch cuts in the complex plane. A recent computation of $D(p^2)$
over the entire complex $p^2$ plane using the Schwinger-Dyson equations (SDE)
\cite{Strauss:2012dg} does not observe
any complex conjugate poles or other cuts than on the negative Euclidean axis. Of course, in order to solve (numerically) the SDE,
truncations had to be made together with a parametrization of the vertices. All the approximations require independent
confirmation and further studies are needed to arrive at a reliable conclusion.
In \cite{Iritani:2009mp}, the gluon spectral function was also probed using fitting forms for the propagator.

\begin{figure}[h] 
   \centering
   \includegraphics[scale=0.4]{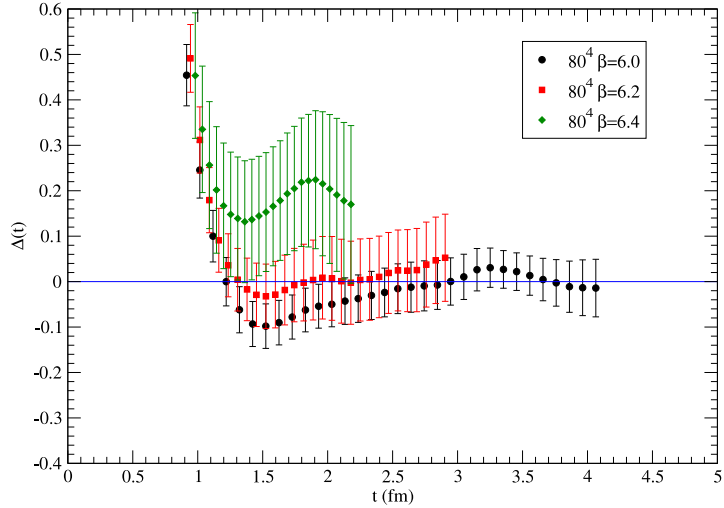}
   \caption{The Schwinger function $\Delta (t)$ for the gluon propagator computed for large lattices and several
                volumes and lattice spacings. Details about statistics and lattices sizes and space and be found in
                \cite{Oliveira:2012eh}.}
   \label{Fig:schwinger_gluon}
\end{figure}

Once the propagator $D(q^2)$ is known, one can compute the Schwinger function defined as
\begin{equation}
   \Delta (t) = \int^{+ \infty}_{- \infty} \frac{dq}{2 \pi} e^{- i q t} D(q^2)  = \int^{+ \infty}_0 dy ~ \rho(y^2) ~ e^{-t y}
   \quad \mbox{ for }Ê \quad t > 0 \ .
\end{equation}
If $\rho ( \mu )$ displays a $\delta$-function singularity, 
then correspondingly $\Delta (t) \sim e^{-m t}$ and $m$ would be
the physical mass of an asymptotic free gluon.
On the other hand, if the Schwinger function has a region where it goes negative, then $\rho (\mu)$ must be negative
over a given region of $\mu$ and, therefore, the gluon cannot be a free particle in the usual sense. In this latter
case we speak about positivity violation by the gluon. This violation of positivity has been
observed both in lattice calculations \cite{Cucchieri:2004mf,Silva2006,Bowman:2007du}
and for the solutions of the Schwinger-Dyson equations \cite{Strauss:2012dg,Alkofer:2003jj}. For completeness, in
Fig. \ref{Fig:schwinger_gluon} we show the Schwinger function computed using the lattice gluon propagator
for several ensembles. Despite the large statistical errors, the plot points towards a $\rho ( \mu)$ with positive
and negative values and one may conclude that the gluon is, certainly, not an asymptotic state contributing to
the S-matrix.

That the gluon is not a physical state can also be viewed from (\ref{Eq:RGZ}). However,
if one combines the gluon propagator of (\ref{Eq:RGZ}) to build a glueball propagator \cite{Dudal:2010cd},
at least part of the glueball propagator looks like a physical particle propagator and it turns out that its mass value is within the
values predicted by quenched lattice QCD \cite{Morningstar:1999rf,Chen:2005mg}.
 In future work, we shall investigate the gluon spectral representation using lattice data, based on the hereafter to be explained technology.

\section{On the glueball propagator}

In this section, we will report on an ongoing computation of the spectral density for the color
singlet glueball operator with quantum number $J^{PC} = 0^{++}$. In the continuum, such
states are associated with $\mathcal{O} = F^2_{\mu\nu}$. A lattice version of $F_{\mu\nu}$ can be found in
\cite{Chen:2005mg}. Our lattice operator for the non-Abelian tensor is such that it reproduces the
continuum $F_{\mu\nu}$ up to corrections of order $a^2$, where $a$ is the lattice spacing.

The lattice simulations were performed using the Wilson action at $\beta = 6.2$, meaning a lattice
spacing $a = 0.0726$ fm as measured from the string tension, and a lattice of $64^4$ with a
physical size of $\sim 4.65$ fm.  We have generated about 872 gauge configurations.

\begin{figure}[h] 
   \centering
   \includegraphics[scale=0.26]{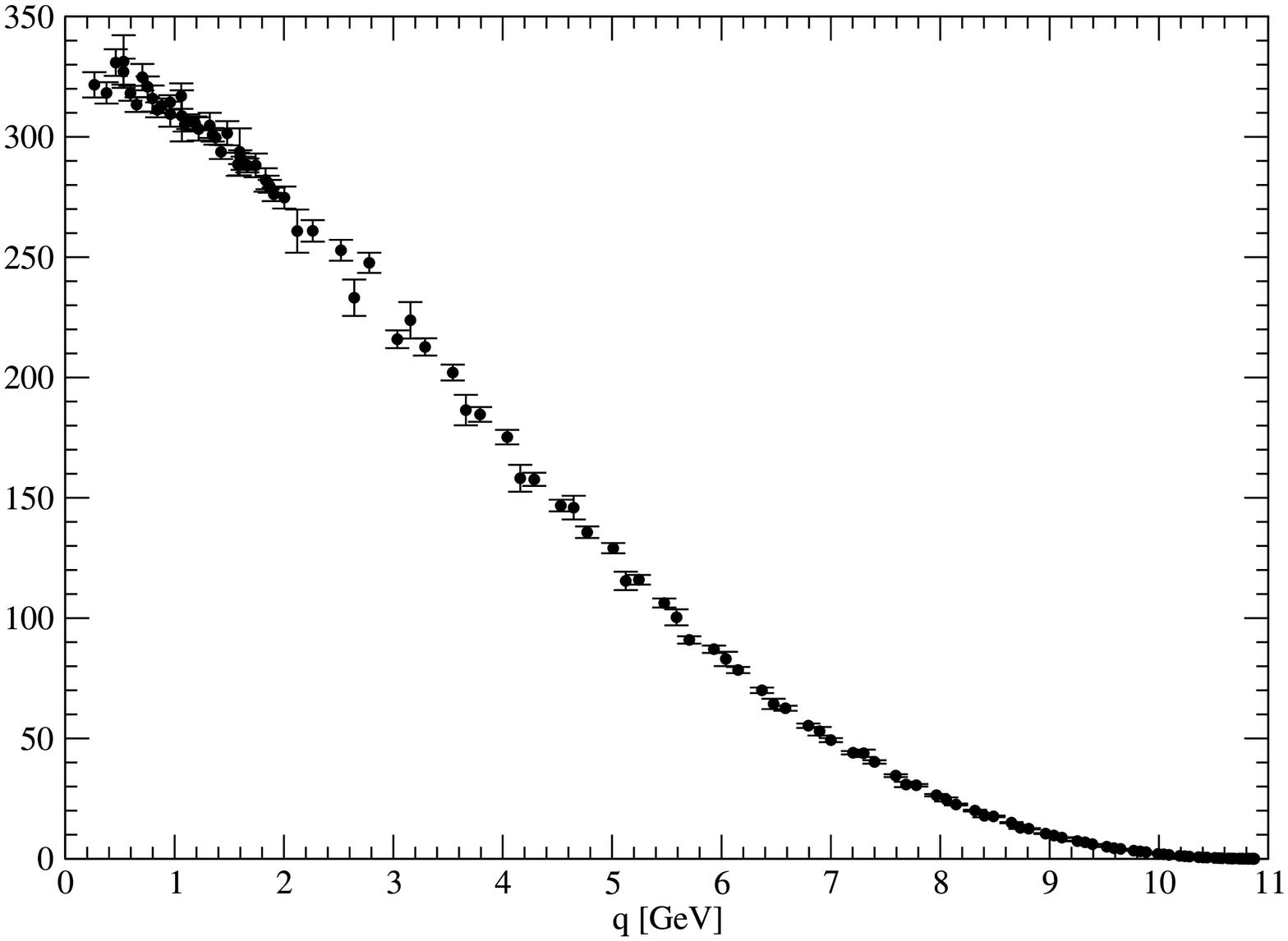}
   \includegraphics[scale=0.26]{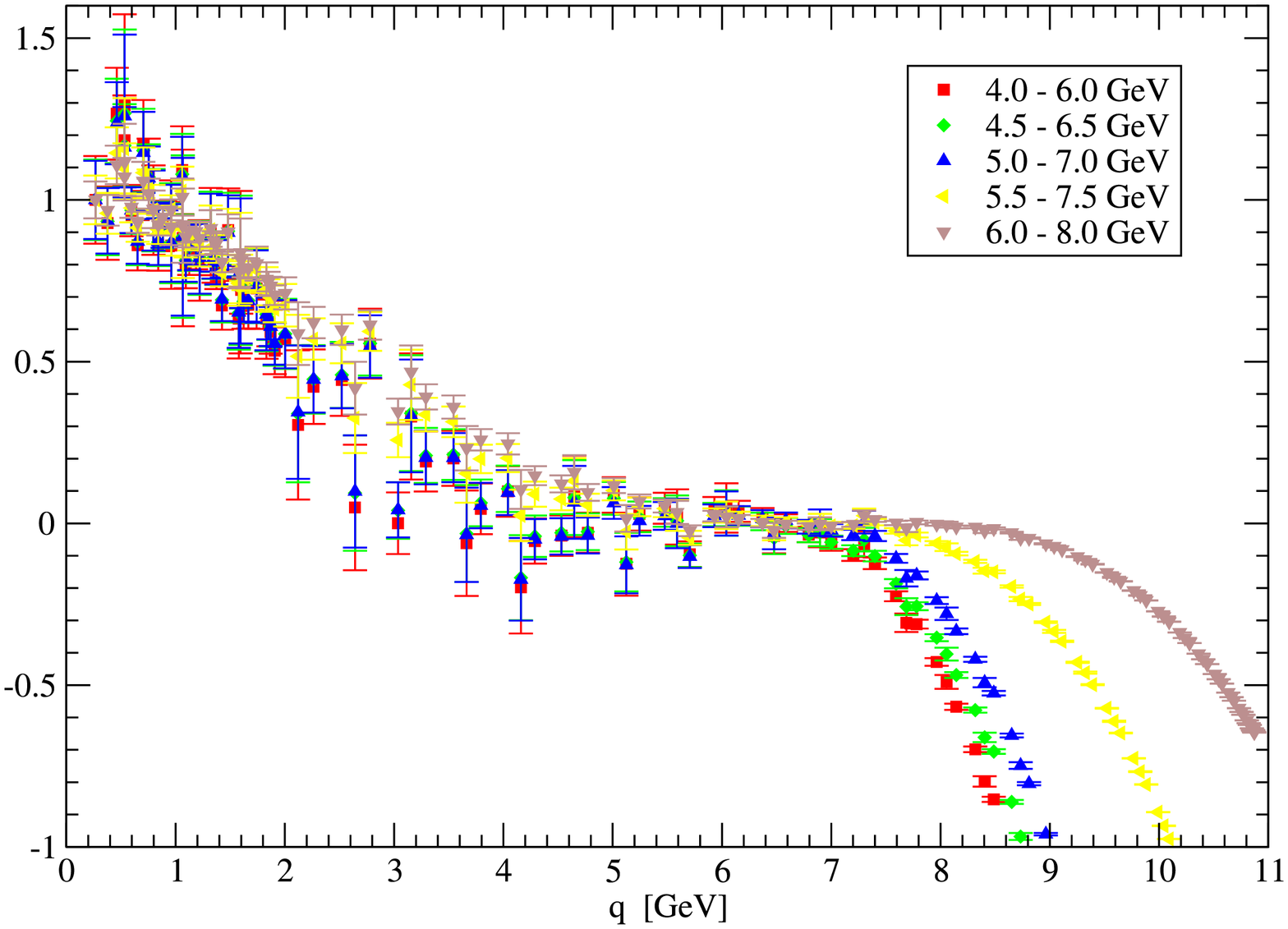}
   \caption{Glueball propagator: (left) bare propagator before subtraction; (right) rescaled subtracted propagator for
                 various fitting intervals. See text for details.}
   \label{Fig:glueball_prop}
\end{figure}

For the glueball propagator one expects 
\begin{equation}
\langle \mathcal{O}(q) \mathcal{O}(-q) \rangle ~ = ~ \mbox{ infinities polynomial in } q^2  ~  + ~
\langle \mathcal{O}(q) \mathcal{O}(-q) \rangle_{\mbox{finite}}  = \int_0^{+\infty} d \mu \frac{ \rho ( \mu ) }{q^2 + \mu}
\ .
\end{equation}
In order to compute the spectral density, one first has to remove the polynomial infinities.
Dimensional analysis shows that the infinities should show up in the coefficients of a polynomial
$\mathcal{P}_{inf} (q^2)$ of order $q^4$ \cite{Dudal:2010cd}.
Therefore, to subtract $\mathcal{P}_{inf} (q^2)$ we fit the bare glueball propagator to
$a_0 + a_1 q^2 + a_4 q^4$ over several momenta interval between 4 and 8 GeV and afterwards
we subtract $\mathcal{P}_{inf} (q^2)$ from the bare propagator.
For the polynomial fit, in all cases, the reduced $\chi^2/d.o.f. \approx 1$.
In Fig. \ref{Fig:glueball_prop} we report on the bare glueball propagator before subtraction (left). The r.h.s.~plot shows
the subtracted and rescaled glueball propagator defined in such a way that $D(q_{min}) = 1$, where $q_{min}$ is
the smallest nonvanishing momenta.

We would like to call the reader's attention that no data points are shown for $q = 0$ in Fig. \ref{Fig:glueball_prop}.
The numerical simulations all give $D(0)$ very large and this point was thence removed from the plot.
In the continuum, we indeed have that
\begin{equation}
 D(0) = \int^{+ \infty}_0 d \mu ~  \frac{\rho ( \mu  )}{\mu} \quad \longrightarrow \quad + \infty
\end{equation}
since the $\rho(\mu)/\mu$ integral also diverges, in good agreement with the discussion of the previous paragraph.

The subtracted and rescaled propagator plot show relative independence of the subtraction point $T$, defined as the
square of the centre of the fitting interval,
 indicative that eventual physical results will be relatively $T$-independent.
This independence of $T$ allows us to
take, from now on, only one of the data sets after subtraction. In the following, we will only consider
the subtracted data obtained from taking the polynomial fit to the range $5 - 6$ GeV.

The finite part, i.e. the glueball propagator after subtraction of $\mathcal{P}_{inf} (q^2)$, is given in terms of
the spectral density as
\begin{eqnarray}
 D(q^2) = \langle \mathcal{O}(q) \mathcal{O}(-q) \rangle_{\mbox{finite}} & = & -
 \left( q^2 - T \right)^3 ~ \int^{+ \infty}_0 d \mu ~ \frac{\rho ( \mu )}{\left( \mu + T \right)^3} ~ \frac{1}{q^2 + \mu}
 \nonumber \\
 & =  &
 - \left( q^2 - T \right)^3 ~ \int^{+ \infty}_0 d \mu ~ \frac{\hat{\rho} ( \mu )}{q^2 + \mu} \ .
\end{eqnarray}
Formally, this equation can be written as follows
\begin{equation}
 \mathcal{D} (q^2) = - \frac{D(q^2)}{\left( q^2 - T \right)^3} = \int^{+ \infty}_0 d \mu ~ \frac{\hat{\rho} ( \mu )}{q^2 + \mu}\equiv\mathcal{L}^2\hat\rho (q^2) \, ,
\end{equation}
i.e.~the l.h.s.~of the equation is given by the double Laplace transform of $\hat{\rho} ( \mu )$, where the Laplace transform reads $\mathcal{L}f(t)=\int_0^{+\infty} e^{-st}f(s)$. The propagator is generated by the Monte Carlo technique and, therefore, we end up with
noisy data as usual. The direct solution of the linear system obtained from the last equation is meaningless: it needs to be regularized. Inverting the Laplace transform is a typical example of an ill-posed problem.

\begin{figure}[h] 
   \centering
   \includegraphics[scale=.5]{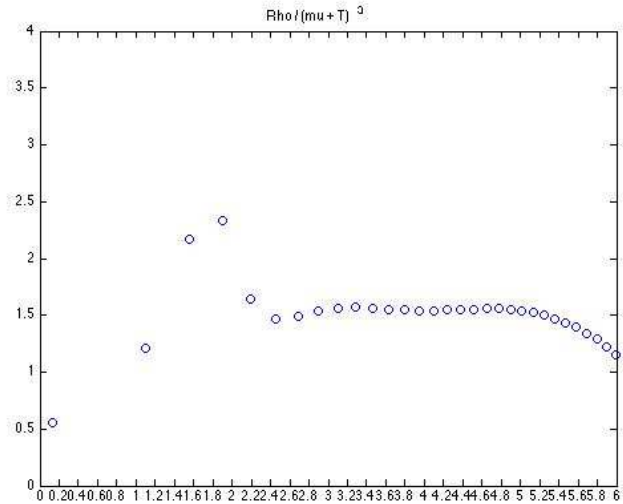}
   \caption{$\hat{\rho}$ as a function of the momentum.}
   \label{Fig:rho_hat}
\end{figure}

We replaced the original linear system by a Tikhonov-Morozov regularized \cite{Tikhonov} linear system,
where the so-called normal equation to be solved is
\begin{equation}
 \mathcal{L}^4\hat\rho(z)+\lambda\hat\rho(z)=\mathcal{L}^2\mathcal{D} (z) \, ,
 \label{Eq:laplace0}
\end{equation}
viz.~
\begin{equation}
 \int^{+ \infty}_{0}dt ~ \hat{\rho}(t) \, \frac{ \ln \frac{z}{t} }{z - t} ~ + ~ \lambda \, \hat{\rho} (z) =
 \int^{+ \infty}_0 dt ~ \frac{\mathcal{D} (t)}{z + t} \, ,
 \label{Eq:laplace}
\end{equation}
where $\lambda$ is the regularization parameter. For solving this equation, 
the (convergent) integrals were numerically handled via the introduction of a cutoff at $\Lambda = 11$ GeV and the system was discretized
via a half-open Newton-Cotes quadrature rule. The solution of (\ref{Eq:laplace}) as a function of
the momentum is show in Fig. \ref{Fig:rho_hat}.

The function $\hat{\rho} ( \mu )$ shows a peak around $\sim 1.7$ GeV and second smaller and broader
peak just above 3 GeV.These preliminary results compare well the predictions of \cite{Morningstar:1999rf,Chen:2005mg}. There, (continuum extrapolated) $0^{++}$ states with masses of 1710(50)(80) MeV and 2670(180)(130) MeV were reported.

\section*{Acknowledgements}
D. D. is supported by the Research-Foundation Flanders (FWO Vlaanderen). 
P. J. Silva acknowledges support by FCT under contract SFRH/BPD/40998/2007. 
O. O. and P.J.S.
acknowledges financial support from the F.C.T. research project PTDC/FIS/ 100968/2008, developed
under the initiative QREN financed by the UE/FEDER through the Programme COMPETE -
Programa Operacional Factores de Competitividade. Part of the computation was performed on the HPC
clusters of the University of Coimbra.
Part of the computational resources (Stevin Supercomputer Infrastructure) and services used in this work were provided by Ghent University, the Hercules Foundation and the Flemish Government - department EWI.
We are grateful for technical support from the ICT Department of Ghent University.



\begin{thebibliography}{99}

\bibitem{MILC}
This work was in part based on the MILC collaboration's public lattice gauge theory code. See
http://physics.utah.edu/$\sim$detar/milc.html


\bibitem{Asakawa:2000tr}
M.~Asakawa, T.~Hatsuda and Y.~Nakahara, Prog.\ Part.\ Nucl.\ Phys.\  {\bf 46}, 459 (2001).

\bibitem{Dudal:2012zx}
D.~Dudal, O.~Oliveira and J.~Rodriguez-Quintero, Phys. Rev. \textbf{D}, to appear [arXiv:1207.5118 [hep-ph]].

\bibitem{Oliveira:2012eh}
O.~Oliveira and P.~J.~Silva, arXiv:1207.3029 [hep-lat].

\bibitem{Cucchieri:2011ig}
A.~Cucchieri, D.~Dudal, T.~Mendes and N.~Vandersickel, Phys.\ Rev.\ D {\bf 85}, 094513 (2012).

\bibitem{Dudal:2010tf}
D.~Dudal, O.~Oliveira and N.~Vandersickel, Phys.\ Rev.\ D {\bf 81}, 074505 (2010).

\bibitem{Bogolubsky:2009dc}
I.~L.~Bogolubsky, E.~M.~Ilgenfritz, M.~Muller-Preussker and A.~Sternbeck, Phys.\ Lett.\ B {\bf 676}, 69 (2009)

\bibitem{Cucchieri:2007md}
A.~Cucchieri and T.~Mendes,  PoS LAT {\bf 2007}, 297 (2007).

\bibitem{Dudal:2008sp}
D.~Dudal, J.~A.~Gracey, S.~P.~Sorella, N.~Vandersickel and H.~Verschelde, Phys.\ Rev.\ D {\bf 78}, 065047 (2008).

\bibitem{Strauss:2012dg}
S.~Strauss, C.~S.~Fischer and C.~Kellermann, arXiv:1208.6239 [hep-ph].

\bibitem{Iritani:2009mp}
 T.~Iritani, H.~Suganuma and H.~Iida, Phys.\ Rev.\ D {\bf 80}, 114505 (2009).

\bibitem{Cucchieri:2004mf}
A.~Cucchieri, T.~Mendes and A.~R.~Taurines, Phys.\ Rev.\ D {\bf 71}, 051902 (2005).

\bibitem{Silva2006}
P. J. Silva, O. Oliveira, Proceedings of Science (LAT2006) 075, arXiv:hep-lat/0609069

\bibitem{Bowman:2007du}
P.~O.~Bowman, U.~M.~Heller, D.~B.~Leinweber, M.~B.~Parappilly, A.~Sternbeck, L.~von Smekal, A.~G.~Williams and J.~-b.~Zhang,  Phys.\ Rev.\ D {\bf 76}, 094505 (2007).

\bibitem{Alkofer:2003jj}
R.~Alkofer, W.~Detmold, C.~S.~Fischer and P.~Maris, Phys.\ Rev.\ D {\bf 70}, 014014 (2004).

\bibitem{Dudal:2010cd}
D.~Dudal, M.~S.~Guimaraes and S.~P.~Sorella, Phys.\ Rev.\ Lett.\  {\bf 106}, 062003 (2011).

\bibitem{Morningstar:1999rf}
 C.~J.~Morningstar and M.~J.~Peardon, Phys.\ Rev.\ D {\bf 60}, 034509 (1999).

\bibitem{Chen:2005mg}
Y.~Chen, A.~Alexandru, S.~J.~Dong, T.~Draper, I.~Horvath, F.~X.~Lee, K.~F.~Liu and N.~Mathur {\it et al.}, Phys.\ Rev.\ D {\bf 73}, 014516 (2006).

\bibitem{Tikhonov}
 See, for example, \emph{An Introduction to the Mathematical Theory of Inverse Problems (A.~Kirsch, Springer, 1996)} and references therein.

\end{thebibliography}
\end{document}